# Increasing ventilation reduces SARS-CoV-2 airborne transmission in schools: a retrospective cohort study in Italy's Marche region

Luca Ricolfi[a,b], Luca Stabile[c], Lidia Morawska[d], Giorgio Buonanno[c,d,*],

[a] Department of Psychology, University of Turin, Italy
[b] David Hume Foundation, Turin, Italy
[c] Department of Civil and Mechanical Engineering, University of Cassino and Southern Lazio, Cassino, FR, Italy
[d] International Laboratory for Air Quality and Health, Queensland University of Technology, Brisbane, Queensland, Australia

**\*Corresponding author:**
Prof. Giorgio Buonanno
Dep. of Civil and Mechanical Engineering – University of Cassino and Southern Lazio
Via Di Biasio 43, 03043 Cassino (FR) – Italy
Email: buonanno@unicas.it

**Abstract**

*Background:* While increasing the ventilation rate is an important measure to remove inhalable virus-laden respiratory particles and lower the risk of infection, direct validation in schools with population-based studies is far from definitive.

*Methods:* We investigated the strength of association between ventilation and SARS-CoV-2 transmission reported among the students of Italy's Marche region in more than 10,000 classrooms, of which 316 were equipped with mechanical ventilation. We used ordinary and logistic regression models to explore the relative risk associated with the exposure of students in classrooms.

*Findings:* For classrooms equipped with mechanical ventilation systems, the relative risk of infection decreased with the increase in ventilation: ventilation ranging from 10 to 14 L s$^{-1}$ student$^{-1}$ reduced the likelihood of infection for students by 80% compared with a classroom with only natural ventilation. From the regression analysis, as confirmed by the predictive theoretical approach, we obtained a relative risk reduction in the range 12%-15% for each additional unit of ventilation rate per person.

*Interpretation:* We need high ventilation rates (> 10 L s$^{-1}$ student$^{-1}$) to protect students in classrooms from airborne transmission; this is higher than the rate needed to ensure indoor air quality. The excellent agreement between the results from the retrospective cohort study and the outcomes of the predictive theoretical approach makes it possible to assess the risk of airborne transmission for any indoor environment.

*Funding:* This research received no specific grant from any funding agency in the public, commercial, or not-for-profit sectors.

# 1 Introduction

A number of studies have investigated transmission routes of respiratory diseases, but few have examined the direct impact of ventilation on indoor transmission (1,2). The SARS-CoV-1 outbreaks (3) in 2004, the MERS-CoV outbreaks (4) and the current SARS-CoV-2 pandemic (5–12) have given a new impetus to research in this field, leading to new evidence and raising awareness of the importance of ventilation and indoor air quality for public health.

Schools represent a critical indoor environment due to the high crowding indexes (number of people relative to the size of the confined space), the long exposure times, and the possible inadequate clean (pathogen-free) air supply. In particular, some studies reported that schools do not amplify SARS-CoV-2 transmission, but rather reflect the level of transmission in the community (13–17). Nonetheless, several SARS-CoV-2 outbreaks in classrooms have been recognized worldwide (5,6), and the situation has worsened with the Omicron variant, which is documented to spread amongst adolescents and children even faster than previous variants of concern (7–9).

As one of the non-pharmaceutical interventions, the government of Italy's Marche region supported the installation of mechanical ventilation systems (MVSs) in approximately 3% of the schools in the region. The objective of this retrospective cohort study was to investigate, through standardized methods for exposure assessment and statistical analysis, the strength of association between ventilation and SARS-CoV-2 airborne transmission in classrooms. We also conducted stratification by including two different sub-cohorts with the aim of evaluating whether the current requirements are sufficient to mitigate the airborne transmission of SARS-CoV-2.

# 2 Materials and methods

## 2.1 Study design and participants

In March 2021, the government of central Italy's Marche region launched a 9 M€ call to fund the installation of MVSs in classrooms to prevent the airborne transmission of SARS-CoV-2 and limit the adoption of distance learning solutions. The funds enabled the installation of mechanical ventilation systems in 316 classrooms (in 56 schools applying for the funding). The population involved in this study consisted of 205,347 students at different educational stages (pre-school 14.6%, primary schools 33.1%, middle schools 18.9%, and high schools 33.4%) attending classes between 13 September 2021 and 31 January 2022. There were 1,419 schools in total included in the study, of which 56 were equipped with an MVS, for a total of 10,441 classrooms with an average occupancy of 20 students per classroom. A total of 10,125 classrooms relied on natural ventilation (i.e. ventilation due to leakage of the building and manual opening of the windows), while 316 were equipped with MVSs.

Infections were investigated in terms of clusters of cases that occurred rather than individual cases of infection; also in accordance with the Italian regulation that defined a cluster as the simultaneous presence in classrooms of 2 positive cases until December 2021 and 3 cases starting from January 2022 (18,19). Temporal exposure was extrapolated from the regional weekly COVID-19 incidence and the relative risk reduction was correlated with the presence of the MVSs in the classrooms. The data was collected by the epidemiological observatory and by the school and infrastructure departments of the Marche region. The David Hume Foundation, a research institution specializing in data analysis, received from the Marche Region the data on the number of positive students in each class for 12 separate sub-periods from September 2021 to January 2022 to identify the clusters as defined above. The entire cohort is represented by the students in the classrooms equipped with MVSs: during the observation period, protective measures were adopted in Italian schools for students such as distancing, use of personal protective equipment (masks), and frequent opening of windows and doors to improve ventilation.

The maximum (nominal) air flow rates of the MVSs installed in the classrooms ranged from 100 to 1,000 $m^3\ h^{-1}$ (with $25^{th}$, $50^{th}$, and $75^{th}$ percentiles equal to 360 $m^3\ h^{-1}$, 600 $m^3\ h^{-1}$, and 800 $m^3\ h^{-1}$, respectively) resulting in a ventilation rate per person ($Q_p$) between 1.4 and 14 L $s^{-1}$ student$^{-1}$ for a classroom with an occupation density of 20 students and with a representative volume of 150 $m^3$, as per the European survey (20). For the purposes of indoor air quality, an air change per hour (ACH) up to 5 h$^{-1}$ is required in Italy (21), corresponding to a $Q_p$ of 10 L s$^{-1}$ student$^{-1}$ for the above-mentioned occupation density and volume. The representative $Q_p$ in European schools ranges from 1.5 to 9 L s$^{-1}$ student$^{-1}$, with lower rates being more representative for natural ventilation (22). Consequently, to stratify the analysis, we also introduced two sub-cohorts: i) sub-cohort 1 represents classrooms with MVSs characterized by a $Q_p$ between 1.4 and 10 L s$^{-1}$ student$^{-1}$ that meet the standard requirements of indoor air quality, also in relation to students' performance (23), and ii) sub-cohort 2 includes classrooms with a $Q_p > 10$ L s$^{-1}$ student$^{-1}$ and up to 14 L s$^{-1}$ student$^{-1}$ and could represent health-based ventilation to protect from airborne transmission.

**2.2 Statistical analysis**

We used simple descriptive statistics to characterize the study population, exposure and risk reduction factors, summarizing quantitative data as means and categorical data as proportions.

Data on the number of positive students are provided as: i) incidence cases (IC), i.e. the number of positive students (provided separately for classrooms with and without MVSs and for different sub-periods); ii) incidence proportions (IP), i.e. the number of positive students per 1,000 students (provided separately for classrooms with and without MVSs and for different sub-periods); and iii) incidence proportion ratio (IPR), i.e. the ratio between the incidence proportion in classrooms with and without MVS.

The risk reduction factors considered in the statistical analysis are: i) the relative risk (RR), i.e. the outcome rate in the classrooms equipped with MVSs divided by outcome rate in the control group (i.e. classrooms without MVSs); ii) the relative risk reduction (RRR), defined as 1-RR, i.e. the proportional reduction of the events in the control group with respect to the investigated one (classrooms with MVS).

To assess the effect of the mechanical ventilation systems on risk reduction we adopted four indicators: i) the cardinal indicator $y_1$ counts the total number of cases in each classroom, subtracts 1 (presumed primary case), and divides the result by the number of students; ii) the cardinal indicator $y_2$ which is similar to $y_1$, except that for classrooms with 5 or more cases, only 4 secondary cases are always counted; iii) the cardinal indicator $y_3$ which is the arithmetic mean between $y_1$ and $y_2$; and iv) the dummy indicator $d_1$ which assumes a value of 1 if a cluster was identified in a classroom.

We developed several ordinary least squares and logistic regression models, including the confounding variables (educational stage and number of students per class) to estimate the net effect of the MVS. Details are reported in the Supplementary material. The data analysis was performed using IBM SPSS Statistics 28.0 and the results are presented as relative risks and 95% confidence intervals (CIs). We used the $\chi^2$ test and Fisher's exact test to compare proportions and the F-test and *t*-test for the statistical significance of the impact of the MVS.

**2.3 Role of the funding source**

The government of the Marche Region as funder of the study had no role in the study design, data organization, data analysis, data interpretation, or writing of the report. LR and GB had full access to the data transmitted by the Marche Region and all the authors had final responsibility for the decision to submit for publication.

**3 Results**

During the entire observation period we recorded 3,121 SARS-CoV-2 infected students (i.e. cases) in 1,004 classrooms, 31 in classrooms equipped with MVSs and 3,090 in classrooms without MVSs (Table 1). The monthly IP, expressed per 1,000 students, was not constant: from 13 September to 23 December 2021 it was lower than 7–31 January, as was the population of the Marche region (Figure 1). Indeed, the IPR for the entire period was equal to 0.32, but it was lower in the period of 7–31 January 2022 (IPR = 0.23) characterized by higher regional incidence cases (IC > 10,000 daily cases) than in the period of 13 September to 23 December 2021 (IPR = 0.45).

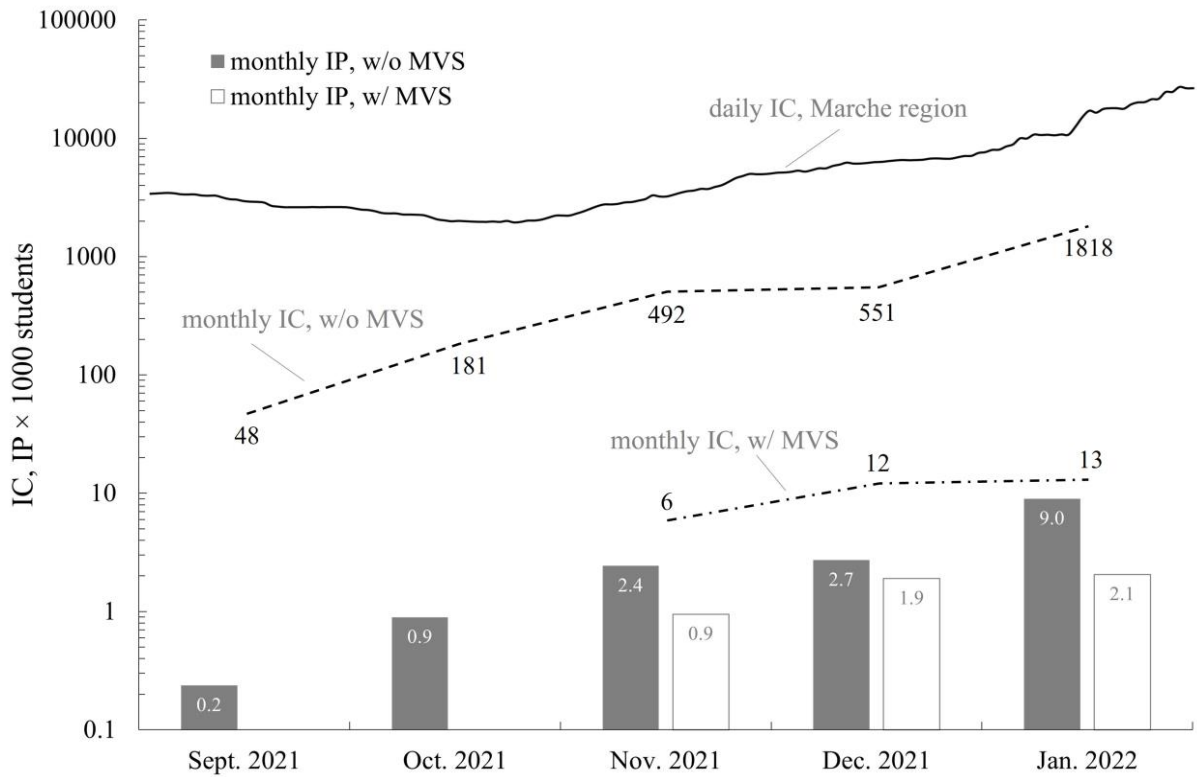

Figure 1. Daily cases in the Marche region, monthly incidence cases (IC) and incidence proportion (IP, per 1000 students) in classrooms with (w/) and without (w/o) mechanical ventilation systems (MVSs) from September 2021 to January 2022.

Table 1. Incidence cases (ICs), incidence proportions (IPs), and incidence proportion ratios (IPRs) observed in classrooms with and without mechanical ventilation systems (MVSs) during the periods of investigation.

| Parameter | Period of investigation | Classrooms without MVS | Classrooms with MVS |
|---|---|---|---|
| Incidence cases, IC | 13 Sept – 23 Dec 2021 | 1,272 | 18 |
| | 7–31 Jan 2022 | 1,818 | 13 |
| | Entire period | 3,090 | 31 |
| Incidence proportion, IP (per 1000 students) | 13 Sept – 23 Dec 2021 | 6.3 | 2.8 |
| | 7–31 Jan 2022 | 9.0 | 2.1 |
| | Entire period | 15.3 | 4.9 |
| Incidence proportion ratio, IPR | 13 Sept – 23 Dec 2021 | 0.45 | |
| | 7–31 Jan 2022 | 0.23 | |
| | Entire period | 0.32 | |

For each indicator, not only did the classrooms equipped with MVS present a lower relative risk (RR) for the entire cohort, but there were also RR and RRR differences in the sub-cohorts according to the level of mechanical ventilation installed (Table 2). Even adopting the most conservative indicator ($y_2$), the results showed that students in classrooms without mechanical ventilation had a 5-fold higher risk of infection compared with sub-cohort 2, and a 3-fold higher risk of infection compared with sub-cohort 1.

Table 2. Relative risks (RRs) and relative risk reductions (RRRs) for the four indicators for the different cohorts.

| | Entire cohort | Sub-cohort 1 | Sub-cohort 2 |
|---|---|---|---|
| RR($y_1$) | 0.19 | 0.21 | 0.15 |
| RRR($y_1$) | 0.81 | 0.79 | 0.85 |
| RR($y_2$) | 0.26 | 0.29 | 0.20 |
| RRR($y_2$) | 0.74 | 0.71 | 0.80 |
| RR($y_3$) | 0.23 | 0.25 | 0.17 |
| RRR($y_3$) | 0.78 | 0.75 | 0.83 |
| RR($d_1$) | 0.09 | 0.13 | 0.0 |
| RRR($d_1$) | 0.91 | 0.87 | 1.0 |

The association between ventilation and infection risk is significant regardless of the location, educational stages, and occupancy as clearly demonstrated in Table 3 where the classrooms were classified in 11 subsamples distinguished by provinces (4 modalities), educational stages (4 modalities), and number of students in the class (3 modalities). The RRRs for the most conservative indicator ($y_2$) are always positive (except in the case of Pesaro, dummy indicator, where -0.33 indicates an increase of the RR in respect to classrooms without MVSs): higher RRRs (even larger than 0.80) were detected in two provinces (Ancona and Macerata), in pre-schools (where no cases were detected, then resulting in null relative risk values), high schools, and in classrooms with more students.

Table 3. Relative risk reduction (RRR) ratios for four indicators and 11 subsamples. RRRs are not provided in cases of null relative risk values.

| | | RRR($y_1$) | RRR($y_2$) | RRR($y_3$) | RRR($d_1$) |
|---|---|---|---|---|---|
| Province | Ancona | 0.91 | 0.88 | 0.89 | - |
| | Ascoli Piceno | 0.67 | 0.52 | 0.60 | - |
| | Macerata | 0.88 | 0.84 | 0.86 | - |
| | Pesaro | 0.41 | 0.29 | 0.33 | -0.33 |
| Educational stage | Pre-schools | - | - | - | - |
| | Primary schools | 0.80 | 0.71 | 0.76 | - |
| | Middle schools | 0.77 | 0.71 | 0.74 | 0.74 |
| | High schools | 0.84 | 0.80 | 0.82 | - |
| Number of students in the classroom | Small classrooms | 0.63 | 0.50 | 0.57 | - |
| | Medium classrooms | 0.83 | 0.77 | 0.80 | 0.83 |
| | Large classrooms | 0.85 | 0.78 | 0.81 | - |

## 4 Discussion

The impact of SARS-CoV-2 on global public health, societies, and economies has been overwhelming. Various containment and mitigation strategies have been implemented to adequately contain the pandemic, with the intention of deferring major surges of patients in hospitals and protecting the most vulnerable people from infection, including elderly people and those with comorbidities. Strategies to achieve these goals are diverse and commonly based on national risk assessments. As well as intense immunization campaigns, public health institutions recommended non-pharmaceutical interventions such as regular cleaning of frequently-touched surfaces, hand washing, universal masking, and social distancing (24), but at the same time they did not convincingly support and propose to increase indoor ventilation to contain infections, mainly due to the dogma on the modalities of transmission of respiratory infections (25). Indeed, even if SARS-CoV-2 transmission can, in principle, occur by spray, inhalation, and touch (26,27), there is an unmistakable increasing body of evidence (28–31) supporting the airborne transmission of SARS-CoV-2 and suggesting that it is primarily an airborne disease transmitted through infectious respiratory fluids released as particles of different sizes and quantities during exhalation (32).

The outcomes of this retrospective cohort study demonstrate a lower incidence of COVID-19 cases in classrooms equipped with MVSs compared with classrooms with natural ventilation, with an IPR of 0.32 over the entire observation period and the entire cohort. The protection from contagion was even greater during the month of January 2022 (0.23), in the presence of high incidence at regional level (> 10,000 cases per day). This outcome suggests that the adoption of MVSs is even more noticeable and effective in periods (or with variants of concern) characterized by high virus circulation. This result was expected because of the key role ventilation plays in reducing occupational hazards according to the engineering level controls described in the traditional infection control hierarchy[37].

One of the major findings of the retrospective study is the estimate of the mechanical ventilation impact on RR. To this end, in addition to the entire cohort we also considered two sub-cohorts: sub-cohort 1 (classrooms equipped with MVSs providing a $Q_p$ up to 10 L s$^{-1}$ student$^{-1}$) and sub-cohort 2 (classrooms with MVSs providing a $Q_p$ greater than 10 L s$^{-1}$ student$^{-1}$). The distinction is important because it allows classrooms to be distinguished between those that comply with Italian law in terms of indoor air quality (sub-cohort 1) and those in sub-cohort 2 which, having ventilation greater than that required by Italian law, could represent the future conditions of health-based protection to control not only thermal comfort, odors, perceived air quality, and energy use, but also respiratory infections. A health-based $Q_p$ between 10 and 14 L s$^{-1}$ student$^{-1}$ in classrooms reduces the likelihood of infection for students in the most conservative case by 80% compared with a classroom relying only upon natural ventilation. The likelihood of infection is reduced by just over 65% in the case of classrooms equipped with mechanical ventilation aimed at indoor air quality requirements. It is therefore evident that ventilation needs to be pushed beyond 10 L s$^{-1}$ student$^{-1}$ (i.e. ACH > 5 h$^{-1}$ for a classroom of 150 m$^3$ with a density of 20 students) in view of ensuring adequate protection from respiratory infectious agents such as SARS-CoV-2. We did find that the impact of the MVSs is greater than that estimated by the raw data (for all the indicators). As an example, if the regression models (details in the Supplementary material) obtained in the case of mechanical ventilation of 10 L s$^{-1}$ student$^{-1}$ are applied, and if we use the more conservative estimates, the average empirical RRR is 0.75, while the corrected value is 0.82 (corresponding to a RR of 0.18). This means that, once the confounding factors (educational stage and number of students per classroom) have been eliminated, the mechanical ventilation is even more incisive than it appeared from an empirical comparison between classes with and without mechanical ventilation.

With reference to eq. (S3) of the Supplementary material and to the indicator $y_2$ (chosen conservatively among the cardinal indicators), in the case of the entire cohort, we obtained a value for RRR per each additional unit of $Q_p$ of 0.15. In the case of a special logistic regression, with outcome variable $y_2$ dichotomized (eq. S5 of the Supplementary material), it was 0.12.

The findings provided by the present retrospective cohort study are extremely important as they confirm the in-field effectiveness of ventilation in terms of risk reduction. Nonetheless, it would be desirable to provide prospective estimates of infection risk in different indoor environments. Recently, we developed a predictive theoretical approach that can estimate the SARS-CoV-2 risk of infection of susceptible individuals via the airborne transmission route when exposed to virus-laden particles emitted by an infected subject in an indoor environment (33–35). The risk of infection is estimated starting from the viral emission rate of the infected subject, the consequent viral concentration in the environment, the resulting viral dose of the exposed susceptible subject and, finally, the adoption of a proper dose–response model to allow the risk to be calculated. The novel aspect of the approach is the a priori evaluation of the viral emission of the infected subject on the basis of the viral load, the expiratory flow rate (influenced by the activity level), and the particle volume concentration expelled by the infectious person (affected by the expiratory activity, i.e. speaking, breathing, etc.). Major details of this predictive approach are reported in our previous papers and are not repeated here for the sake of brevity (33,34,36–38). Figure 2 shows the comparison between the RRs observed in the investigated classrooms and those estimated through the theoretical predictive approach for a specific scenario as a function of the ventilation rate per person. The scenario reported here considers: viral load and infectious dose typical of the Delta variant of concern (this variant was prevalent during the study period (36,39,40)), an average classroom volume of 150

m$^3$, an infected student breathing only for the entire school time (5 hours), exposed subjects performing only sitting/standing activities, and the effectiveness of masks on the reduction of the risk ranging from 0% (no mask) to 80% (actual reduction for respirators)[44]. As expected, the simulated RR values decrease as a function of the ventilation rate per person with quite similar results to the observed RR for sub-cohort 2 (i.e. > 10 L s$^{-1}$ student$^{-1}$): the RR obtained from the simulation at 14 L s$^{-1}$ student$^{-1}$ was equal to 0.24 ± 0.04.

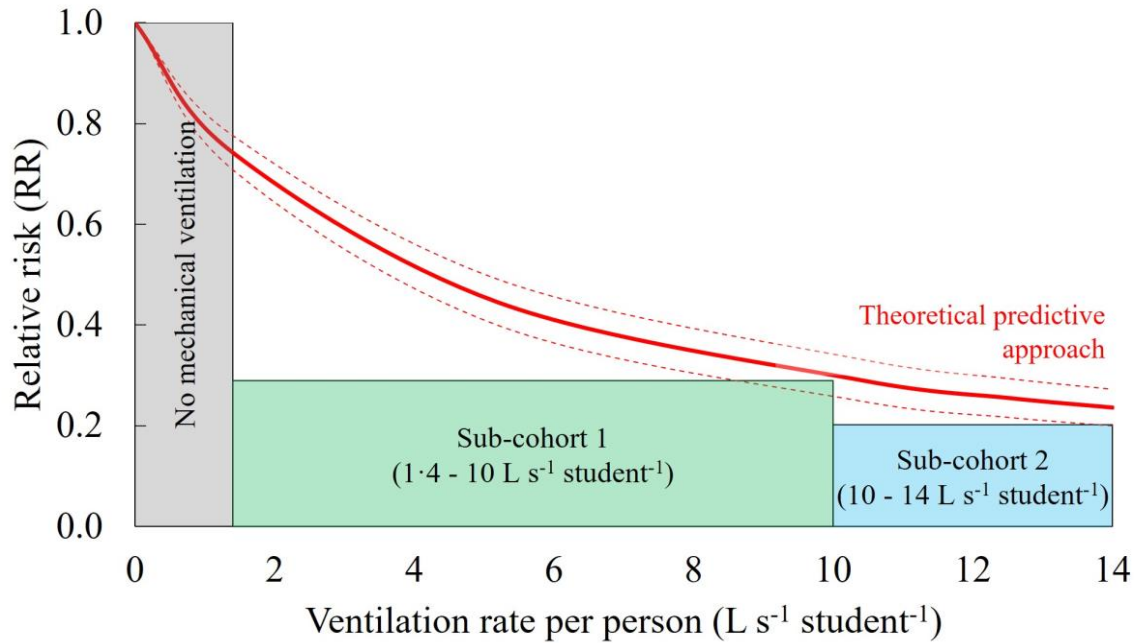

Figure 2. Comparison between the relative risk (RR) observed in the investigated classrooms and the RR estimated through the theoretical predictive approach for a specific scenario as a function of the ventilation rate per person, $Q_p$.

The agreement between the results obtained from the retrospective cohort study and the values calculated through the predictive approach (33) is the second important finding of this paper. Indeed, this result represents a further validation of the approach through a retrospective cohort study that follows the experimental validation that we recently performed through an experimental study conducted under controlled conditions in a hospital room (41). Such validations confirm the possibility of extending the use of the approach, once the scenario has been defined, to any indoor environment of interest in addition to school classrooms and providing predictive estimates of the effectiveness of ventilation for different exposure scenarios and variants of concern.

Organizations such as the American Society of Heating, Refrigerating and Air-Conditioning Engineers and the Federation of European Heating, Ventilation and Air Conditioning Associations have ventilation standards, but they must be improved to explicitly consider infection control in addition to thermal comfort, odor control, perceived air quality, initial investment cost, energy use, and other performance issues in the management of indoor environments. Indeed, despite national regulations, the technical standards do not consider infection control when suggesting design air change per hour in indoor environments. As an example, the European standard EN-16798 (42,43) suggests an $Q_p$ in classrooms of > 10 L s$^{-1}$ student$^{-1}$ only for occupants with special needs (children, elderly, persons with disabilities, etc.; referred to as Category I) and for "non-low-polluted buildings" (i.e. buildings where no effort has been made to select low-emitting materials and where activities with emission of pollutants are not limited or prohibited).

Some limitations of this study should be noted. Firstly, further studies would provide deeper knowledge about the reduction of contagion risk in schools as a function of ventilation; for example, classrooms equipped with high ventilation should be investigated to identify a possible limit threshold beyond which the benefits in terms of risk reduction become negligible. Secondly, we considered a generalized operation of controlled mechanical ventilation systems at maximum flow rate. However, this was a reasonable hypothesis during the emergency period, which coincided with the observational period. Thirdly, our study was limited to SARS-CoV-2: other respiratory pathogens would require different ventilation rates per person.

To the best of our knowledge, this is the largest retrospective cohort study in schools aimed at assessing the impact of mechanical ventilation in mitigating the risk of COVID-19 infection. The results demonstrate the effectiveness of adequate ventilation ($> 10$ L s$^{-1}$ student$^{-1}$) and the possibility of applying these techniques in a similar way in all indoor environments that represent the natural habitat of humans and which require complex, targeted management, not only of the control of thermal comfort, odors, perceived air quality, and energy use, but also of respiratory infections (44).

**Contributors**
LR and GB conceived the study. LR, GB, LS, and LM contributed to the study design. LR contributed to the planning of statistical methods and data analysis. LR was involved in data collection, training of the study team, and data analysis. GB was the primary author of the manuscript. LM, LR, and LS contributed as senior authors on the manuscript draft. LR, LM, and LS commented on the manuscript draft. All authors have seen and approved the submitted manuscript.

**Declaration of interests**
We declare no competing interests.

**Acknowledgments**
The authors thank the Marche Region for the availability in the transmission of the incidence data to the David Hume Foundation relating to the classrooms of the region, of which 316 have been equipped with mechanical ventilation systems through several calls for a total of 9 M € .

**Supplementary Material**

The indicators used, three cardinal ($y_1$, $y_2$, $y_3$) and one dichotomous ($d_1$), show positive and statistically significant correlations as reported in the correlation matrix of Table S 1 (sequence of eigenvalues: 2.732, 0.202, 0.066).

Table S 1. Correlation matrix amongst the selected indicators.

|       | $y_1$ | $y_2$ | $y_3$ | $d_1$ |
|-------|-------|-------|-------|-------|
| $y_1$ | 1     | 0.91  | 0.98  | 0.80  |
| $y_2$ |       | 1     | 0.97  | 0.89  |
| $y_3$ |       |       | 1     | 0.86  |
| $d_1$ |       |       |       | 1     |

The relationship between the mechanical ventilation system (MVS) and each of the four indicators was always statistically significant and such that the MVS always reduces the transmission of the virus. Table S 2 shows the results of the *t* test of equality of the means (for independent samples) for the three cardinal indicators ($y_1$, $y_2$, $y_3$). Table S 3 summarizes the results of the significance tests for the impact of the MVS on the dummy $d_1$.

Table S 2. Results of the *t* test of equality of the means for the cardinal indicators ($y_1$, $y_2$, $y_3$).

| Indicator | *t* | *p*-value (unilateral) |
|-----------|-------|------------------------|
| $y_1$     | 2.765 | 0.006                  |
| $y_2$     | 2.981 | 0.003                  |
| $y_3$     | 2.930 | 0.003                  |

Table S 3. Results of the statistical tests for the dummy indicator ($d_1$).

| Test | *p*-value (bilateral) | *p*-value (unilateral) |
|------|-----------------------|------------------------|
| Chi square | 0.001 | - |
| Fisher | <0.001 | <0.001 |
| Odds ratio (Confidence interval 95%) | 0.01–0.56 | |

In addition to mechanical ventilation, we have identified two additional factors (confounding parameters) that affect airborne transmission: the educational stage and the number of students per class. For these two factors, a dummy was introduced that distinguishes between compulsory school (CS, elementary and middle schools) and other educational stages, and a term of interaction between the number of students (NS) in the class and the absence of mechanical ventilation systems (1-MVS). These two variables have been added to the mechanical ventilation to estimate the parameters of an ordinary regression model with a cardinal dependent variable (eq. S1) and of a logistic regression model with a dichotomous dependent variable (where y = 1 means that mechanical ventilation influences the airborne transmission, eq. S2).

$$y = b_0 + b_1 \cdot ACH + b_2 \cdot CS + b_3 \cdot NS \cdot (1 - MVS) + error \quad (S1)$$

$$\frac{p(y=1)}{1-p(y=1)} = e^{\left(b_0 + b_1 \cdot ACH + b_2 \cdot CS + b_3 \cdot NS \cdot (1-MVS)\right)} \quad (S2)$$

The regressions were performed for all the indicators (3 cardinal, 1 dummy), obviously using the ordinary regression for the dependent variables $y_1$, $y_2$, $y_3$, and the logistic regression for the dependent dummy variable $d_1$. The effect of the MVS is always negative and statistically significant.

Table S 4. Main results of the regressions conducted for all four indicators (using the ordinary regression for the cardinal indicators $y_1$, $y_2$, $y_3$ and the logistic regression for the dummy indicator $d_1$) and the confounding parameters: $b_1$

refers to mechanical air change per hour (ACH), $b_2$ to compulsory schools (CS), and $b_3$ to the number of students in the classroom (NS).

| Dependent variable | Constant $b_0$ | | ACH (h$^{-1}$) $b_1$ | | CS (-) $b_2$ | | NS (1-MVS) (-) $b_3$ | |
|---|---|---|---|---|---|---|---|---|
| | Coeff | Sig | Coeff | Sig | Coeff | Sig | Coeff | Sig |
| Ordinary regressions | | | | | | | | |
| $y_1$ | 0.670 | 0.000 | -0.171 | 0.010 | 0.656 | 0.000 | 0.035 | 0.002 |
| $y_2$ | 0.674 | 0.000 | -0.157 | 0.005 | 0.648 | 0.000 | 0.031 | 0.001 |
| $y_3$ | 0.672 | 0.000 | -0.161 | 0.006 | 0.652 | 0.000 | 0.033 | 0.001 |
| Logistic regression | | | | | | | | |
| $d_1$ | -3.535 | 0.000 | -0.625 | 0.014 | 0.655 | 0.000 | 0.080 | 0.000 |

**Relative risk estimate: direct approach**

To quantify the effect of ventilation on airborne transmission, we compared the empirical relative risk (RR) per mechanical ventilation rate per person ($Q_p$) with that obtained from a special logistic regression model. The mechanical $Q_p$ can be related to the air change per hour (ACH) of the classroom by considering a classroom occupation density of 20 students per classroom and a classroom volume of 150 m$^3$.

Given two values of unit of mechanical $Q_p$ ($Q_{p1}$ and $Q_{p2}$) and the corresponding estimated RR, we calculated the relative risk reduction (RRR) per unit $Q_p$, (1-α), using the relationship:

$$RR = \alpha^{(Q_{p2}-Q_{p1})} \Rightarrow \alpha = RR^{1/(Q_{p2}-Q_{p1})} \tag{S3}$$

Applying this equation to the entire cohort we obtained α = 0.85, corresponding to an RRR for each additional unit of ventilation rate per person of 15%.

**RR estimate: logistic regression approach**

The estimate of the RRR per unit ventilation rate per person can be also obtained by applying logistic regression to the conservative indicator $y_2$ once it undergoes a dichotomization process. The dichotomization applied here is: the dummy $y_2$ assumes the value of 1 with the same percentage of the observed cases. The α value can be obtained as follows:

$$\alpha = e^{b_1} \tag{S4}$$

The corresponding α value is 0.88, corresponding to an RRR for each additional unit of ventilation rate per person of 12%.

Table S 5. Main results of the logistic regressions conducted for the dummy variable $y_2$ (after dichotomization) and the confounding parameters: $b_1$ refers to mechanical air change per hour (ACH), $b_2$ to compulsory schools (CS), and $b_3$ to the number of students in the classroom (NS).

| Dependent variable | Constant $b_0$ | | ACH (h$^{-1}$) $b_1$ | | CS (-) $b_2$ | | NS (1-MVS) (-) $b_3$ | |
|---|---|---|---|---|---|---|---|---|
| | Coeff | Sig | Coeff | Sig | Coeff | Sig | Coeff | Sig |
| Logistic regression | | | | | | | | |
| Dummy ($y_2$) | -3.276 | 0.000 | -0.133 | 0.011 | 0.746 | <0.001 | 0.033 | 0.002 |